\documentclass[article]{aa}
\usepackage{graphicx}
\usepackage{aalongtable}
\usepackage{lscape}
\usepackage{txfonts}
\usepackage{natbib}
\bibpunct{(}{)}{;}{a}{}{,}
\usepackage{color}
\newcommand{\figref}[1]{Fig. \ref{#1}}
\newcommand{\mymag}{\hbox{$\,.\!\!^{\rm{m}}$}}
\newcommand{\myarcsec}{\hbox{$.\!\!^{\prime\prime}$}}
\newcommand{\myarcmin}{\hbox{$.\!\!^{\prime}$}}
\newcommand{\myarcsecnodot}{\hbox{$\;\!\!^{\prime\prime}\;$}}
\newcommand{\myarcminnodot}{\hbox{$^{\prime}\;$}}

\newcommand{\angstrom}{\AA}
\newcommand{\angstromblank}{\AA$\;$}
\def\EE#1{\times 10^{#1}}

\def\cm3{\rm ~cm^{-3}}
\def\kms{\rm ~km~s^{-1}}

\def\funit{\rm ~erg~s^{-1} cm^{-2} sr^{-1}}

\def\Msun{~{\rm M}_\odot}

\def\lsim{\!\!\!\phantom{\le}\smash{\buildrel{}\over

  {\lower2.5dd\hbox{$\buildrel{\lower2dd\hbox{$\displaystyle<$}}\over

                               \sim$}}}\,\,}

\def\gsim{\!\!\!\phantom{\ge}\smash{\buildrel{}\over

  {\lower2.5dd\hbox{$\buildrel{\lower2dd\hbox{$\displaystyle>$}}\over

                               \sim$}}}\,\,}

\begin{document}
\title{Observational and theoretical constraints for an H$\alpha$-halo
     around the Crab Nebula}
\subtitle{
\thanks{Based on observations made with ESO Telescopes at the La Silla
     and Paranal Observatories, Chile (ESO Programmes 66.D-0489, 68.D-0096 and
     170.A-0519).}$^,$\thanks{Based on observations made with the Nordic
     Optical Telescope, operated on the island of La Palma jointly by Denmark,
     Finland, Iceland, Norway, and Sweden, in the Spanish Observatorio del
     Roque de los Muchachos of the Instituto de Astrofisica de Canarias}}

   \author{ A. Tziamtzis\inst{1} \and
 M. Schirmer\inst{2,3} \and
 P. Lundqvist\inst{1} \and
 J. Sollerman\inst{1}}

   \institute{ Department of Astronomy, AlbaNova University Center, Stockholm University, 106 91 Stockholm, Sweden \and
    Isaac Newton Group of Telescopes, Calle Alvarez Abreu 68 2, E-38700 Santa Cruz de La Palma, Spain  \and
    Argelander-Institut f\"ur Astronomie, Universit\"at Bonn, Auf dem H\"ugel 71, 53121 Bonn, Germany}

   \offprints{Anestis Tziamtzis, \email{anestis@astro.su.se}}
   \date{\today}

%\abstract{}{}{}{}{}
% 5 {} token are mandatory

\abstract
{} 
% context heading (optional) % leave it empty if  necessary
{We searched for a fast moving H$\alpha$ shell around the Crab nebula. 
  Such a shell could account for this supernova remnant's missing mass, and
  carry enough kinetic energy to make SN 1054 a normal Type II event.}
{Deep H$\alpha$ images were obtained with WFI at the 2.2m MPG/ESO telescope
  and with MOSCA at the 2.56m NOT. The data are compared with theoretical
  expectations derived from shell models with ballistic gas motion, constant
  temperature, constant degree of ionisation and a power law for the density
  profile.} 
{We reach a surface brightness limit of $5\EE{-8}\funit$. A halo is detected,
  but at a much higher surface brightness than our models of recombination
  emission and dust scattering predict. Only collisional excitation of
  Ly$\beta$ with partial de-excitation to H$\alpha$ could explain such
  amplitudes. We show that the halo seen is due to PSF scattering and thus not
  related to a real shell. We also investigated the feasibility of a
  spectroscopic detection of high-velocity H$\alpha$ gas towards the centre of
  the Crab nebula. Modelling of the emission spectra shows that such gas
  easily evades detection in the complex spectral environment of the 
  H$\alpha$-line.}
{PSF scattering significantly contaminates our data, preventing a detection of
  the predicted fast shell. A real halo with observed peak flux of about 
  $2\EE{-7}\funit$ could still be accomodated within our error bars, but our 
  models predict a factor 4 lower surface brightness. 8m class telescopes 
  could detect such fluxes unambiguously, provided that a sufficiently 
  accurate PSF model is available. Finally, we note that PSF scattering also
  affects other research areas where faint haloes are searched for around 
  bright and extended targets.}

{}
\titlerunning{Constraints for an H$\alpha$-halo around the Crab SNR}

\authorrunning{A. Tziamtzis et al.}

   \maketitle

%\maketitle

%
%________________________________________________________________

%%%%%%%%%%%%%%%%%%%%%%%%%%%%%%%%%%%%%%%%%%%%%%%%%%%%%%%%%%%%%%%%%%%%%%
%%%%%%%%%%%%%%%%%%%%%%%%%% INTRODUCTION %%%%%%%%%%%%%%%%%%%%%%%%%%%%%%
%%%%%%%%%%%%%%%%%%%%%%%%%%%%%%%%%%%%%%%%%%%%%%%%%%%%%%%%%%%%%%%%%%%%%%

\section{\label{intro}Introduction}
According to historical records, SN 1054 was visible during daytime for more
than three weeks and for almost two years during night-time \citep[see
e.g.][]{cla77}. Its lightcurve \citep[e.g.][]{sll01}, hydrogen-rich nebula, and
the presence of a central neutron star \citep[][]{sta68,com69} are all
compatible with a normal core-collapse event. However, the velocities of the
filaments in the Crab nebula are rather low \citep[$\sim1400 \kms$,][]{wol72},
as is the observed ejecta mass of $4.6\pm1.8\Msun$ \citep{fsh97}.

The elemental abundances and the low velocity and mass of the filaments may be
accomplished by a low-energy explosion of a $8-13 \Msun$ star \citep[][]
{nom85,nom87,kjh06}. Indeed, supernovae with very low kinetic energies are 
known, such as SN 1997D with $(1-4) \times 10^{50}$ erg \citep{tur98,cu00}.
However, no events with as little kinetic energy as SN 1054 ($1.5\EE{49}$
erg) have yet been identified.

\cite{che77,che85} suggested that a hydrogen-rich shell of a few solar 
masses could carry most of the kinetic energy. Such a shell could be observed
as extended optical emission, or as non-thermal radio and thermal X-ray
emission caused by a shock front advancing through the interstellar medium
such as in the case of SNR G21.5-0.9 \citep[in X-ray,][]{mat05}.

Searches for such emission around the Crab nebula have been performed at 
radio wavelengths \citep[][]{wiw82,vel84,vel85,tru86,vel92,fkc95}, in the 
optical \citep[e.g.][]{gul82,fsh85} and in X-rays
\citep[][]{mag85,mag89,prs95,sew06}, yielding negative results. Possibly the
Crab nebula expands into a local bubble of the ISM with correspondingly low 
interaction rates \citep[see][and references therein]{wlk99}.

The first evidence for the existence of a shell was reported by \cite{hes96},
in the form of [O~III] emission surrounding the outer boundaries of the
nebula. \cite{shs98} interpreted this emission as due to radiative shocks at
the outer edge where the synchrotron nebula accelerates the hypothetical
shell. \cite{sll00} detected a blueshifted absorption feature in C~IV
$\lambda$1550 in a far-UV HST spectrum, extending up to $\sim 2500 \kms$. Such
a feature of the shell was predicted by \cite{lfc86} using photoionisation
models. Lower limits of $0.3 \Msun$ and $1.5\EE{49}$ erg were derived for the 
mass of the shell and its kinetic energy, respectively. This is significantly
below the canonical value of $10^{51}$ erg, but the results of \cite{sll00} do 
not rule out a normal Type II explosion. For a comprehensive summary of the
search for a halo and for a general overview of the Crab nebula see
\cite{hes08}.

By means of sufficiently deep H$\alpha$ surface brightness maps strong
constraints can be put on the shell mass and its radial density profile.
\cite{sll00} showed that a fast shell with 4$\Msun$ would escape the deepest
H$\alpha$ search so far, reaching a surface brightness limit of
$1.5\EE{-7}\funit$ \citep[][using spectroscopy]{fsh97}. In this paper
we describe our attempts to image an H$\alpha$ halo with the Wide Field Imager
(WFI) \citep{bmi99} at the 2.2m MPG/ESO telescope and the Mosaic Camera
(MOSCA) at the 2.56m Nordic Optical Telescope, reaching a sensitivity of
$5.0\EE{-8}\funit$. The observations and data reduction are summarised in
Sect. 2. We analyse the images in Sect. 3 and link the results to theoretical
models in Sect. 4. The feasibility of a direct spectroscopic detection towards
the centre of the Crab nebula is evaluated in Sect. 5. Our results are
discussed in Sect. 6, followed by our conclusions. Throughout this paper we
use the term \textit{shell} for actual physical material around the Crab
nebula, and the term \textit{halo} for its appearance in images and spectra.

%%%%%%%%%%%%%%%%%%%%%%%%%%%%%%%%%%%%%%%%%%%%%%%%%%%%%%%%%%%%%%%%%%%%%%
%%%%%%%%%%%%%%%%%%%%%%%%%%%% THE DATA %%%%%%%%%%%%%%%%%%%%%%%%%%%%%%%%
%%%%%%%%%%%%%%%%%%%%%%%%%%%%%%%%%%%%%%%%%%%%%%%%%%%%%%%%%%%%%%%%%%%%%%

\begin{table}
\caption{Summary of the WFI, MOSCA and FORS1 observations}
\label{expdates}
\begin{tabular}{l l l l l}
\hline
\hline
Instrument & Exp. time [sec] & Seeing & Observation\\
 & (number of exp.) & [$^{\prime\prime}$] & dates\\ 
\hline
\noalign{\smallskip}
MOSCA & 20700 (23) & 1\myarcsec0 & 2006-12-20\\
WFI & 23520 (36) & 1\myarcsec1 & 2000-11-06/07\\
 & & & 2002-02-02/05/06\\
\hline
\end{tabular}
\end{table}

\section{Observations and data reduction}
\subsection{\label{obs}Imaging data}
WFI covers 34\myarcminnodot$\times$ 33\myarcminnodot with a pixel scale of
0\myarcsec238. We observed in the H$\alpha$ filter
($\lambda_{c}=6588$~\angstrom, FWHM$=$74~\angstrom) using a very wide dither
pattern (see Fig. \ref{wfi_geom}) in order to suppress instrumental effects
and to obtain a good superflat. This is especially important when searching
for low surface brightness features \citep[see][and in particular their
Fig.~8]{esd05}. Observations were carried out in service mode in dark and grey
time (see Sect. \ref{superflatting} for details).

Independent H$\alpha$ images ($\lambda_c=6564$ \angstrom, FWHM$=$33 \angstrom,
NOT filter code \#21) were taken with MOSCA in dark time. MOSCA has 4 CCDs and
offers in this filter a comparatively small, 5\myarcmin5 diameter circular
field with 0\myarcsec217 per pixel. It was centred on the north-western part
of the Crab nebula (see Fig. \ref{m1neg}).

\subsection{Data Reduction}
We processed the multi-chip camera data with the THELI\footnote{Available at
http://www.astro.uni-bonn.de/$\sim$mischa/theli.html} pipeline \citep{esd05}.
The images were overscan corrected (not for MOSCA, which does not have an
overscan), debiased, flatfielded, gain-corrected, and in case of WFI also
superflatted. The MOSCA data could not be superflatted as the target covered
most of the field of view. One of the MOSCA CCDs (\#4) did not flatfield
properly and was partially masked (see also Fig. \ref{mosca_ha}). The images
were then individually weighted and we determined the relative photometric
zeropoints between them. After sky subtraction and a full astrometric
calibration the images were coadded. The photometric zeropoint is constant
across the stacked mosaic images with an error of $\sim$0\mymag07.

In the following we describe the superflatting and sky subtraction. These two
steps are critical as they must not suppress or enhance a halo while
correcting for flat field residuals and inhomogeneous sky background.

\subsubsection{\label{superflatting}Superflatting}
\begin{figure}[t]
  \includegraphics[width=1.0\hsize]{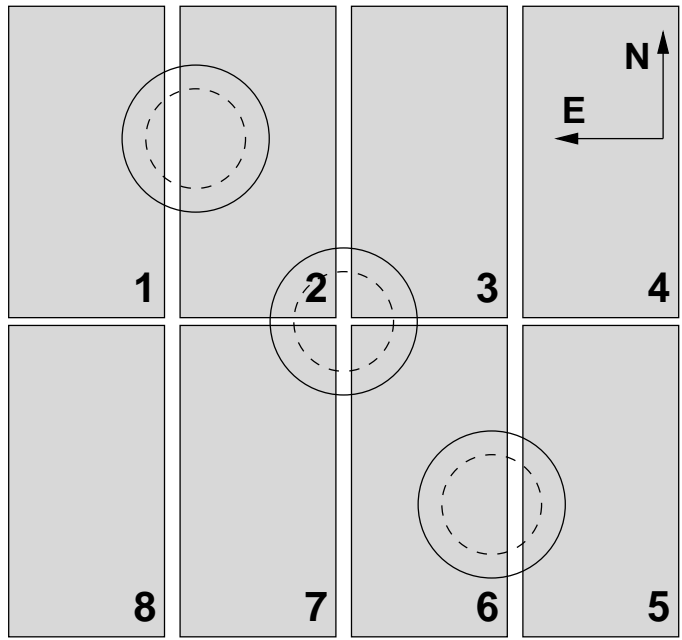}
  \caption{\label{wfi_geom}{The WFI detector layout and the dither pattern. 
      The Crab nebula was placed on three main positions on the detector and 
      is represented to scale by the solid circles. At each of these main 
      positions random dither offsets of up to $\pm 2\myarcmin5$ were
      applied (small dashed circles).}}
\end{figure}

Three main dither positions (\figref{wfi_geom}) were chosen for the WFI
observations. Dither offsets of $4^{\prime}-5^{\prime}$ were made at each one. 
Since the Crab nebula is 6\myarcminnodot in size, this approach ensured
that every pixel saw the sky much more often than the object, which is the
basic requirement when making superflats. We first run \textit{SExtractor}
\citep{bea96} to detect all objects consisting of at least 5 connected pixels,
 each of which having a minimum S/N-ratio of 2. In this manner stars and
the Crab nebula were masked out. Lower detection thresholds would have removed
sky features as well which must be preserved. The masked images of a given CCD
were then scaled to the same mode and median-combined using a 2.5$\sigma$
outlier rejection. This is the standard approach in THELI.

11 of the 36 images were taken in grey conditions, with the moon 12
degrees or less above the horizon and 80 to 110 degrees away from the 
target. 7 of these exposures were significantly affected by uneven sky
background or reflections and removed from further processing. In the 
remaining 4 grey exposures only one or two of the chips were affected 
(none with the Crab nebula inside) and not used for the superflat.

The superflats were then smoothed with a 1\myarcminnodot wide kernel in order
to suppress small scale noise. Since CCDs 1 and 8 still showed significant,
time-variable background features in all exposures, they were excluded at this 
stage.

\subsubsection{Sky subtraction} 
After superflatting and rejecting CCDs 1 and 8, only small ($1\%-3\%$) and
large-scale ($\sim10$\myarcminnodot) gradients were left in the exposures. They
are larger in size than the Crab nebula and a possible halo. The outermost
2\myarcminnodot of the right edges of CCDs 4 and 5 are occasionally brighter
by about 2\%. This is not a cause of concern as the dithering never moved the
target close to this region.

THELI uses a similar approach for background subtraction as for the superflat.
All objects with at least 5 connected pixels and 2$\sigma$ above the sky 
noise were replaced with the mode of the remaining pixels. This intermediate
image was then smoothed and subtracted. We tested two different smoothing
scales of 1\myarcmin8 and 3\myarcmin0, and as an additional check also created
a coadded image where only a constant sky was subtracted from each CCD. A very
similar halo around the Crab nebula is detected in all three coadded images,
thus it was not artificially introduced by the data reduction. In fact, we
show below that it is caused purely by PSF scattering. The coadded images with
the 1\myarcmin8 and 3\myarcmin0 background smoothing scales are very similar,
and we conservatively chose the latter for further analysis. A comparison with
the constant sky subtraction image shows that the modelling lowered the halo's
brightness by less than $\sim10$\%.

The sky background of the MOSCA data was not modelled, as only small areas of
sky were available between the Crab nebula and the edge of the field. We
subtracted constant estimates obtained from empty areas located
$\sim2$\myarcminnodot from the nebula's outer edge. Some particular issues
with the MOSCA reductions are explained in the Appendix.

\begin{figure}[t] 
\center 
\includegraphics[width=1.0\hsize]{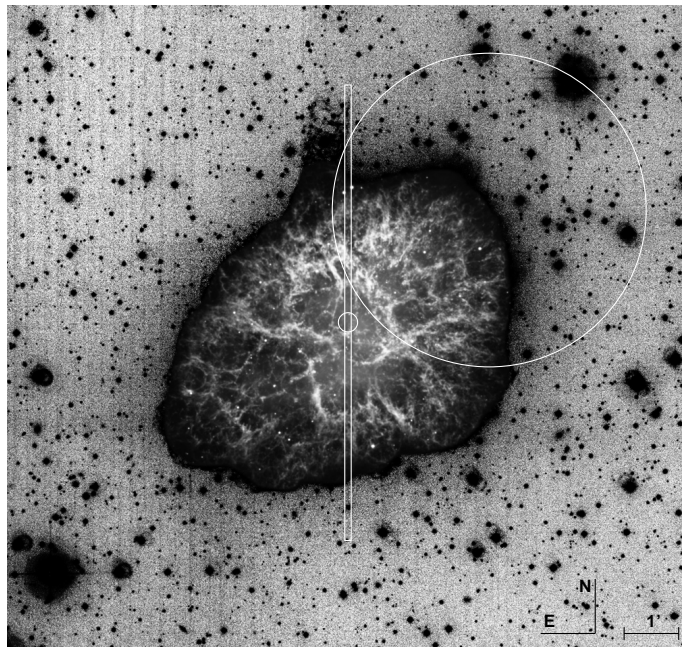} 
\caption{Part of the coadded WFI image. For clarity we did not invert the
  centre of the nebula. A faint halo around the SNR is visible. Superimposed
  is the slit position for the FORS1 spectra, together with the location where
  we tested the feasibilty of a spectroscopic detection of a fast shell. The
  large circle shows the MOSCA field of view.}
\label{m1neg}
\end{figure} 

\subsubsection{Absolute photometric calibration}\label{absphotometry}
We used the Crab pulsar and its flux-calibrated spectrum \citep[from][]{sll00} 
for the absolute photometric calibration of the WFI data. Since the pulsar
lies within a complex nebular environment, we must remove the contribution of
the latter from the pulsar's flux. We obtained a DAOPHOT PSF model from
isolated, bright but unsaturated, field stars and subtracted it from the
pulsar's PSF. In this way the nebular contribution could be quantified and
removed. The pulsar's magnitude was then aperture corrected, using the same
field stars. The differences between aperture and PSF photometry were then
$\leq$ 0.01 mag, and the final photometric zeropoint became 
ZP$_{\rm WFI}=21.42\pm0.15$. 

Flux-calibration for the MOSCA data was based on DAOPHOT aperture 
photometry of the spectro-photometric standard star HD 93521. After
application of the MOSCA filter transmission curve to the reference spectrum,
the photometric zeropoint became ZP$_{\rm MOSCA}=21.5\pm0.10$. With this
calibration there was a constant offset of 0.07 mag in field stars seen by
MOSCA as compared to WFI. The Carlsberg Meridian 
Archive\footnote{http://www.ast.cam.ac.uk/$\sim$dwe/SRF/camc\_extinction.html} 
shows increased extinction levels for the corresponding week in La Palma,
however without a measurement for our night. The observations were thus
probably not made in photometric conditions, and we corrected the MOSCA
zeropoint for this offset.

\begin{figure}[t]
\center 
\includegraphics[width=1.0\hsize]{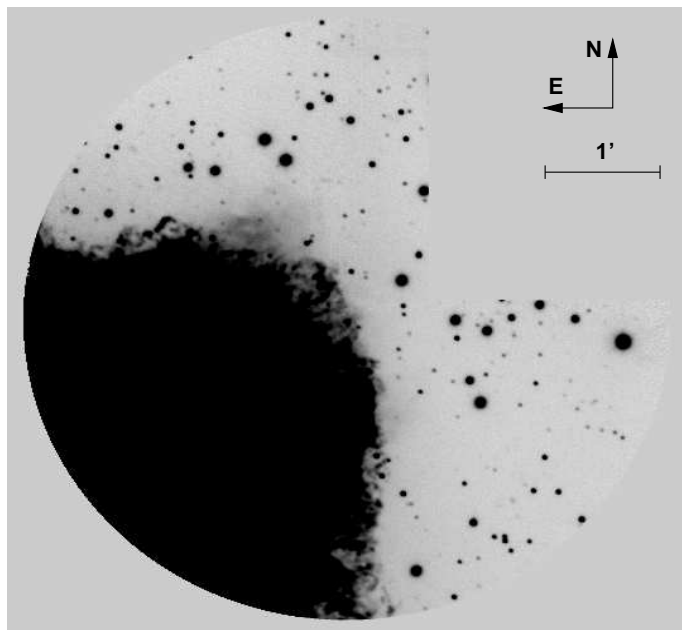} 
\caption{The coadded MOSCA image. The missing part to the upper right is due
  to a partially masked CCD.}
\label{mosca_ha}
\end{figure}

\section{\label{radialprofile}Observational results}
In this section we present the results from our imaging campaigns. We trace 
clear but inconsistent signals in both the WFI and MOSCA data and discuss them
in Sect. \ref{modelpredictions}.

Our halo detection strategy is described in the following based on the WFI
data. It was done in a similar way for MOSCA. We 
\begin{itemize}
\item{mask all stellar flux so that it does not contribute to the halo,}
\item{define an outer boundary of the Crab nebula,}
\item{measure and average the halo brightness as a function of distance from
    this boundary in many different sectors.}
\end{itemize}

\subsection{\label{masking}Masking the image}
In order to mask any unwanted light, we ran \textit{SExtractor} with a 
detection threshold of $2\sigma$ and a minimum number of 15 connected
pixels.  The noise in coadded mosaic images is very uneven due to the
dither pattern and the gaps between the CCDs, and this fact must be taken
into account in the detection process. The coadded weight map reflects these
noise properties and was therefore slotted into \textit{SExtractor} as
well. In this way stars, filter ghosts, and the Crab Nebula itself were
masked while the halo was preserved. Some residuals of scattered light were
still visible around brighter stars and masked manually. These masks are shown
as black regions in \figref{wfi_sectors}. Also shown in this Figure is
the contribution of PSF scattering mimicking a halo, which we discuss in
Sect. \ref{psfscattering}.

\subsection{\label{defineboundary}Defining the boundary around the Crab
  nebula}
\begin{figure}
\center
\includegraphics[width=1.0\hsize]{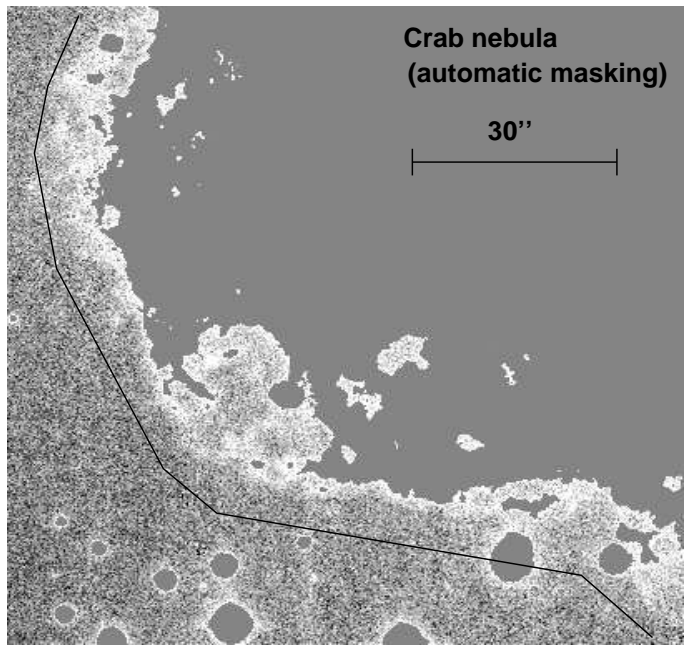}
\caption{Defining the boundary (mask) of the Crab nebula. Only the 
  south-eastern area is shown. The automatic masking with \textit{SExtractor}
  left a fractal-like border which hampers the determination of the starting
  point for the halo measurement in a sector. The solid line shows the manual
  re-definition of the mask. Most of the stellar flux was automatically
  masked, but remaining residuals around brighter stars had to be covered
  manually. These masks can be seen in Fig. \ref{wfi_sectors}.} 
\label{coastline}
\end{figure}

\begin{figure}[t]
\center
\includegraphics[width=1.0\hsize]{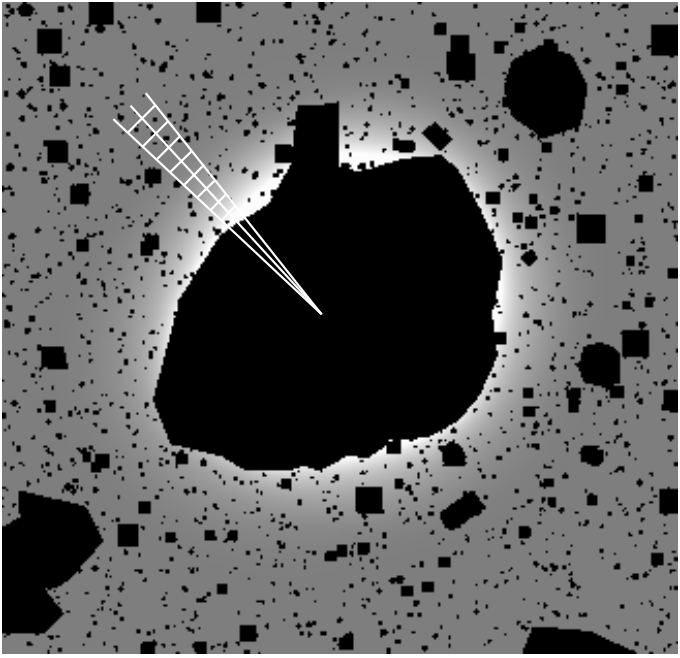}
\caption{Masks for the field stars, together with two of the 72
  measurement sectors. These converge towards the explosion centre and
  are drawn to scale. The density of the radial bins is four times higher than
  indicated, and the field of view is similar to that in Fig. \ref{m1neg}. 
  The brightening in the immediate surrounding of the Crab nebula shows the
  amount of PSF scattering expected for WFI (see Sect. \ref{psfscattering}).}
\label{wfi_sectors}
\end{figure}

We measure the halo brightness as a function of distance from the nebula in
several sectors protruding outwards from the explosion centre. Once the
sectors cross the edge of the nebula, the radial coordinate (distance) is set
to zero, and we begin integrating the halo flux. However, the edge's
fractal-like appearance (\figref{coastline}) makes the determination of the
zero-point of the radial coordinate difficult. Isolated or half-connected
groups of pixels still reside inside, leading to an underestimation of the
distance between the explosion centre and the outer edge. As a result,
the stack of individual profiles is not properly aligned, leading to increased
scatter in the result. To avoid this, we defined the manual mask indicated by
the solid line in \figref{coastline}.
% This new approach reduced the error bars
%for the halo profile by about 20\%, implying that our result is unbiased by
%the nebula's irregular shape.

\subsection{\label{theradialprofile}The radial halo profile}
We assume the centre of the halo to coincide with the explosion centre,
located at $\alpha_{J2000}= 05^h 34^m 32.7^s$, $\delta_{J2000}= +22^{\circ}
00^{\prime} 48\farcs7$ \citep[back-projecting the pulsar proper motion as
given by][]{kcg08}. The exact position is not a critical parameter for our
analysis, the measured halo profile changes negligably when the assumed
explosion centre is shifted by a few arcseconds. We defined 72 sectors, each 5
degrees wide, emerging from this centre. Those sectors that covered the
jet-like Crab chimney \citep[see][for details]{rfy08} were excluded from our
analysis, hence 68 (MOSCA: 20) sectors remain.

In order to scan the profile in each sector, we divided them into 29 (MOSCA:
19) equidistant radial bins 24 pixels in length corresponding to 5\myarcsec7
(5\myarcsec2) for WFI (MOSCA). Their mean brightness was calculated using an
interative 2.5$\sigma$ outlier rejection. This clipping procedure reduced the
error bars of the final radial profile by just 5\%, indicating that our masks
are already very efficient.

\begin{figure}[t]
\center
\includegraphics[width=1.0\hsize]{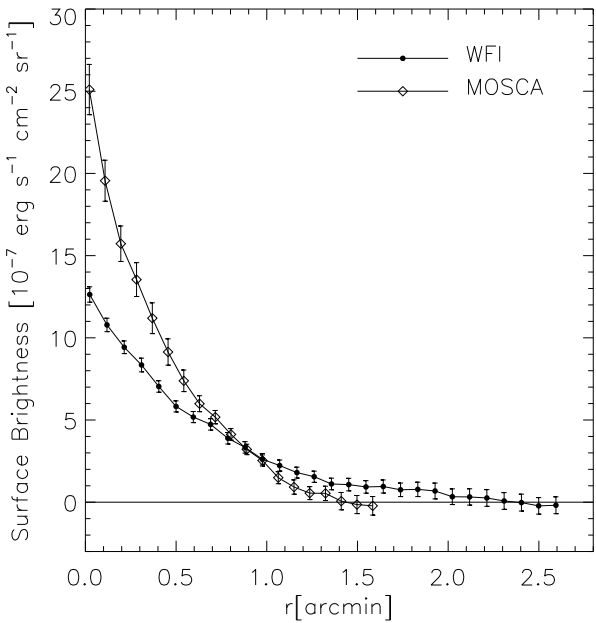}
\caption{The halo profiles observed. They fall off following power laws with
  slopes of $-0.88$ (WFI) and $-1.12$ (MOSCA).}
\label{haloprofile_wfi_mosca}
\end{figure}

Figure~\ref{haloprofile_wfi_mosca} shows the results. The error bars
represent the standard deviation of the mean surface brightness and are
dominated by brightness fluctuations as a function of azimuthal angle. We
can trace the haloes to distances of $1\myarcmin9$ and $1\myarcmin3$ for WFI
and MOSCA, respectively. The MOSCA signal peaks at $2.5\EE{-6} \funit$ and is
twice as high as that for WFI, but it also decays faster. If $x$ is the
distance from the outer edge of the nebula in arcminutes, then the profiles
are proportional to $10^{-0.88 x}$ (WFI) and $10^{-1.12 x}$ (MOSCA). These
inconsistencies are investigated in Sect. \ref{modelpredictions}. The
sensitivity limit reached is $5\EE{-8}\funit$, corresponding to 31.2 mag per
square arcsecond.

\section{\label{modelpredictions}Analysis}
We show here that the haloes are caused by PSF scattering, and we infer upper
limits to the surface brightness of a real shell. We present model
calculations for three different physical processes that can lead to H$\alpha$
emission in the shell and compare them to our observations. 

\subsection{\label{psfscattering}PSF scattering}
\begin{figure}[t]
\center
\includegraphics[width=1.0\hsize]{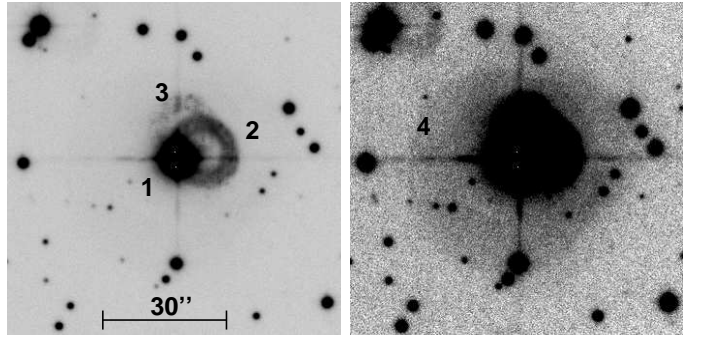}
\caption{Illustrates the WFI filter ghosts. The star shown is the brightest
  seen in the lower left edge of Fig. \ref{m1neg}. The core of the PSF is
  labelled as \textit{1}, followed by three filter ghosts of decreasing
  intensity, labelled $2-4$.}
\label{wfifilterghosts}
\end{figure}

\begin{figure}[t]
\center
\includegraphics[width=1.0\hsize]{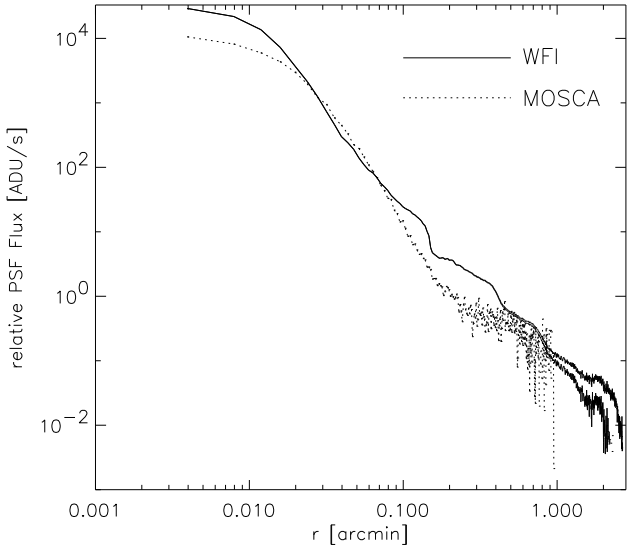}
\caption{The two PSF models for WFI (solid line) and MOSCA (dotted line). The
  bumps in the WFI PSF at 0.15, 0.4 and 0.8 arcmin reflect
  the various filter ghosts. The MOSCA PSF is extrapolated for radii larger
  than 0\myarcmin85 by a power law with index $\beta=-2.4$ (not shown). MOSCA
  has a significantly more compact PSF than WFI for radii larger than about
  4$^{\prime\prime}$ (0\myarcmin06), which explains the steeper slope of the
  halo in that data.}
\label{wfi_mosca_psf}
\end{figure}

Figure \ref{wfifilterghosts} shows filter ghosts and other scattered light
around stars. In order to evaluate how much this effect contributes to the 
haloes, we obtained azimuthally averaged PSF models for both data sets (see 
Fig. \ref{wfi_mosca_psf}).

The PSF model for WFI extends to a radius of 2\myarcmin7 and was created
from very bright stars in the coadded image. The saturated core was replaced
by one obtained from unsaturated stars. During the construction of the model a
strict $1\sigma$ rejection threshold was applied to suppress contamination
from field stars. The resulting PSF profile exhibits several bumps due to the
filter ghosts which make simple mathematical fits inadequate. We also created
a second PSF model with reduced wing amplitude, by subtracting the RMS
value of the sky background on a large scale from the first model. In this
way we could evaluate how sensitive light scattering is to small changes in
the amplitude of the PSF wings.

The PSF model for MOSCA was obtained in a similar way from two moderately
bright stars in a single 600s blank field exposure taken in the same night.
The model's empirical part extends out to 0\myarcmin85, beyond which it is
extrapolated by a power law with index $\beta=-2.4\pm0.4$. The slope was
determined from the profile between 0\myarcmin45 and 0\myarcmin85. Note that
the uncertainty of this model at a radius of 0\myarcmin2 is already larger
than that for the WFI PSF at 2\myarcmin0.

\begin{figure}[t]
\center
\includegraphics[width=1.0\hsize]{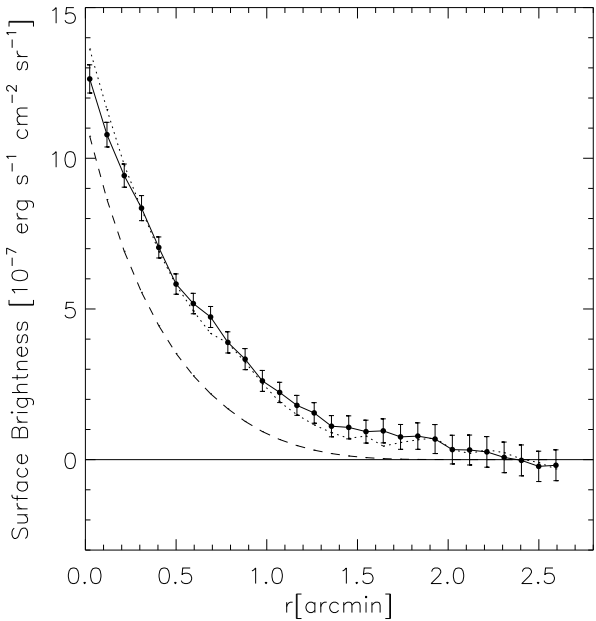}
\caption{The WFI halo profile with the contributions from the PSF (dotted
  line), and from the PSF with reduced wing amplitude (dashed).}
\label{wfi_halo}
\end{figure}

To evaluate the PSF contribution, we zeroed all pixels in the coadded images 
having values less than 5 times the sky noise, keeping only the cores of stars
and the Crab nebula. These images were then convolved with the PSF models, and
the profile measurements were repeated with identical settings and masks. The
result for WFI is shown in Fig. \ref{wfi_halo} (dotted line), together with
the contribution from the PSF model with lowered wing amplitude (dashed
line). The latter significantly underestimates the halo flux as compared to
the unaltered model. Residual light from field stars contributes to the halo
only for radii larger than 1\myarcmin5. We evaluated this by convolving an
image where all field stars were masked out. Within the error bars of the
best-fitting PSF and of the measurement, a real halo with an H$\alpha$ peak
flux of  $2\EE{-7}\funit$ can still be accomodated, comparable to the upper
limit of $1.5\EE{-7}\funit$ obtained by \cite{fsh97}.

\begin{figure}[t]
\center
\includegraphics[width=1.00\hsize]{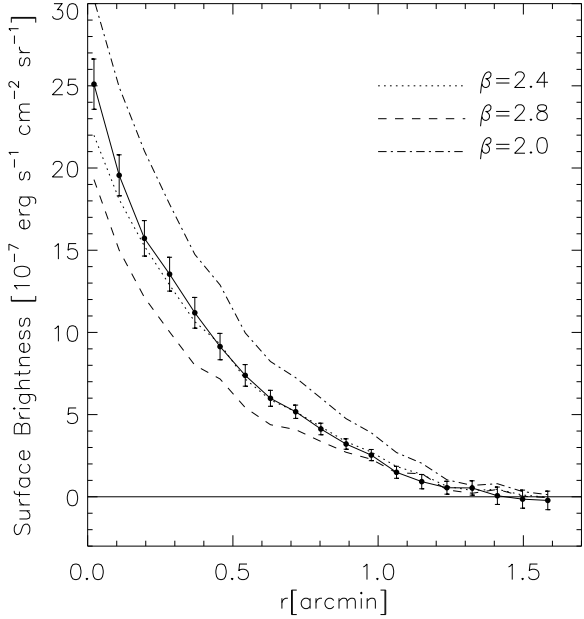}
\caption{The MOSCA halo profile and the best fit PSF contribution (dotted
  line). The other lines show the 1$\sigma$ uncertainty of the fit.}
\label{mosca_halo}
\end{figure}

Figure \ref{mosca_halo} shows the corresponding result for MOSCA. The halo is 
very well explained by PSF scattering (dotted line). The other two lines
represent the $1\sigma$ uncertainty of the power-law fit to the PSF wing.
Due to the less reliable MOSCA PSF model we could not improve the upper flux
limit obtained with WFI.

We stress that simplified approaches such as a comparison of the slope of the 
PSF with the slope of the halo, or of the brightness ratios between inner and 
outer part of the PSF and inner and outer part of the Crab nebula, are 
insufficient. In these ways PSF scattering would have been underestimated by 
about 3 orders of magnitude. Only full 2D convolutions can quantify the effect
accurately. It appears plausible that an earlier detection made in H$\alpha$
on a photographic plate \citep{muc81,mur94} also shows this effect. 
\cite{daf85} argued that the detection made was of instrumental nature, but
the subject was not investigated further.

\subsection{\label{shellmodels}Theoretical models for H$\alpha$ emission from
  a fast shell}
Our data still allows for a peak H$\alpha$ surface brightness of 
$2\EE{-7}\funit$ through the WFI H$\alpha$-filter. The upper limit of
$1.5\EE{-7}\funit$ by \cite{fsh97} refers to the total unfiltered surface 
brightness. We compare these values below with predictions based upon the
shell models of \cite{sll00}. Therein, the shell is of spherical shape and
expands freely outside the Crab nebula with $V\propto R$. It has an inner
minimum radius of $R_{\rm min}=5.0\times10^{18}$ cm corresponding to a minimum
velocity of $1680\kms$. Unlike the Crab nebula the shell is not affected by
the pulsar wind. The mass of the shell is set to $M_{\rm sh} = 4.5 \Msun$, it
has a fixed electron temperature of $T_{\rm e}= 2\EE4$~K \citep{lfc86}, and a
radial density profile $\rho \propto R^{-\eta}$, with $\eta = 3$ or 
$\eta = 4$. Smaller values for $\eta$ do not fit the profile of the C~IV
absorption \citep[see][]{sll00}, and values $\eta\gtrsim5$ would make the
H$\alpha$ surface brightness surpass our detection limit.

All fluxes determined from the models must be corrected for galactic
extinction before they can be compared to observations, using the reddening
parameters $E(B-V)=0.52$ and $R=3.1$ \citep{sll00}. Our dereddened upper limit
of the H$\alpha$ surface brightness through the WFI H$\alpha$-filter 
becomes $6.6\EE{-7}\funit$.

\subsubsection{\label{recombemission}Recombination emission}
\begin{figure}[t]
\center
\includegraphics[width=85mm, angle=0, clip]{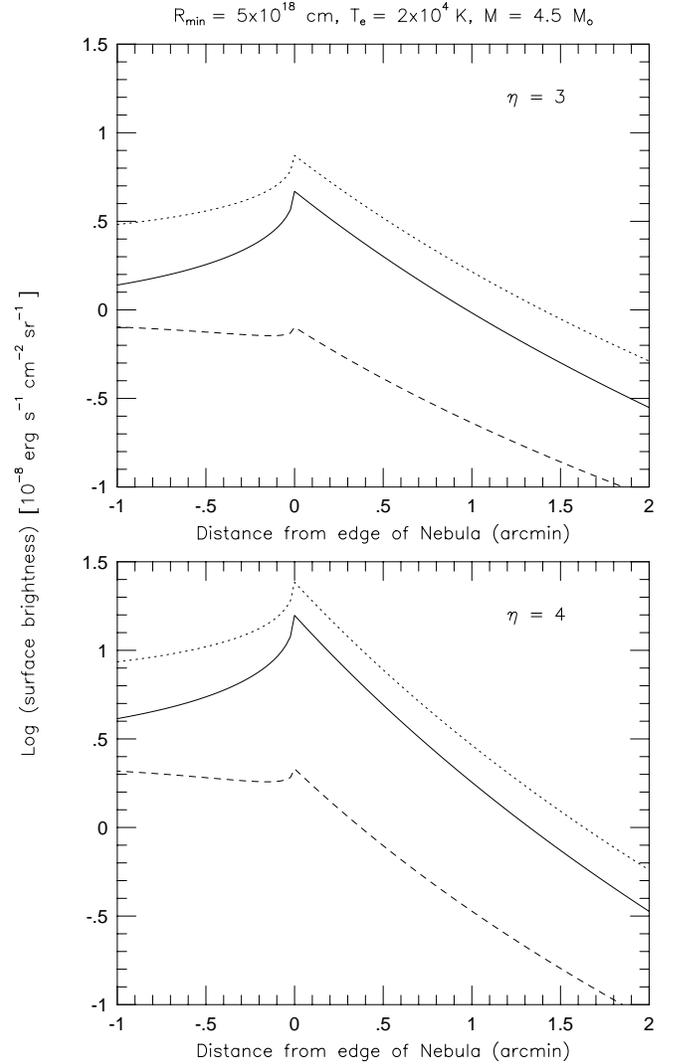}
\caption{H$\alpha$ halo surface brightness using the model described in Sect.
  \ref{shellmodels}. The dotted lines show the unfiltered total H$\alpha$
  surface brightness, whereas the solid and dashed lines show what is captured
  through H$\alpha$ and continuum ESO/WFI filters, respectively. Note how the
  surface brightness falls off rapidly with distance from the edge of the
  nebula. As discussed in the text, the $\eta = 4$ model provides an adequate
  fit to the shape of the fall-off, but it cannot explain the flux amplitude.}
\label{surface_halpha_plot}
\end{figure}

For the modelling of the recombination emission we furthermore assume that 
hydrogen in the shell is fully ionised and that the gas has a He/H-ratio of 
0.1 by number. Figure \ref{surface_halpha_plot} shows the modelled
surface brightness profile as a function of distance from the edge of the
nebula for two cases: $\eta = 3$ (top panel) and $\eta = 4$ (bottom). The
solid lines describe the surface brighness captured by the WFI
H$\alpha$-filter.

For comparison we also plot the signals expected through the WFI red 
continuum filter (dashed line) and for unfiltered observations (dotted
line). The continuum filter selects $H\alpha$-emission from shell gas
receding from the observer with $2000-6000 \kms$. The advantage is
that the Crab nebula is much less bright in these wavelengths and thus the
amount of scattered light is reduced, which favours a detection. On the
other hand, the gas seen in this filter is actually at significantly larger
distance behind the Crab nebula and just projected next to the nebula's 
edge. The densities in these parts of the shell are much smaller and thus 
the emission is significantly reduced (see Fig. \ref{surface_halpha_plot}).
The dotted line represents the fluxes expected for unfiltered observations
which we need later on for comparison with spectroscopy.

Peak surface brightness for the H$\alpha$-filter occurs at the edge of the 
nebula with $1.5\EE{-7}\funit$ for $\eta = 4$. For comparison, the observed 
WFI halo peaks at $1.2\EE{-6}\funit$ which translates into a dereddened value 
of $4.0\EE{-6}\funit$, $\sim 26$ times higher than the prediction. Taking the
PSF scattering into account, the $\eta=4$ (3) model is 4 (14) times below our
dereddened upper limit.  A similar relation emerges for the upper limit 
of $1.5\EE{-7}\funit$ by \cite{fsh97}. Their result is based on unfiltered 
spectra and must therefore be compared with the dotted lines in Fig. 
\ref{surface_halpha_plot} at roughly 0.3 arc min from the edge of the nebula. 
Including reddening, the $\eta = 4$ (3) model is 4 (10) times below their
detection limit. For a temperature of the shell of $7.5\EE3$~K, as assumed by
\cite{fsh97}, the modelled surface brightness is $\sim 2.7$ times higher than
for $2\EE4$~K. This brings the predicted surface brightness closer to the
imaging and spectroscopic upper limits by the same factor. For the $\eta=4$ 
(3) model the limits would only be a factor of $\sim 1.5$ larger than the 
emitted surface brightness. The recombination emission can also be enhanced to
some degree if the gas is clumpy and has a higher gas density. The latter,
however, would quickly exceed a sensible value for the shell mass, make the
C~IV~$\lambda$1550 absorption seen by \cite{sll00} too high, and contradict 
the findings of \cite{shs98}.

The slope of the haloes expected from recombination emission is close to
exponential with $10^{-0.60 x}$ for $\eta = 3$ and $10^{-0.82 x}$ for 
$\eta = 4$. We note that the latter is coincidentally very similar to the 
slope for the WFI halo caused by PSF scattering. The MOSCA data would require
steeper density profiles than $\eta=4$.

\subsubsection{Dust scattering}
Scattering by interstellar dust can contribute to the shell flux, too. The
Crab nebula is an isolated object, and its dust density is nearly equal to the
surrounding ISM, as was found by observations from the 
optical to sub-millimeter wavelengths \citep{fsb90,gre04,tem06}. The latter 
authors also showed that small-grain dust is entirely absent inside the
nebula, and that larger grains amount to only $10^{-3}-10^{-2}\Msun$. Hence a
normal dust-to-gas ratio can be assumed and an analysis similar to that in
\cite[][chapter 7]{ost89} is valid. We adopt an H$\alpha$ luminosity of 
$7\EE{34}$~erg~s$^{-1}$ for the nebula \citep{dav87} and take the electron
density to be $3 \cm3$ close to it, as is appropriate for the $\eta = 4$ model
described above. In this way one obtains a ratio of emissivities (scattered
emission on dust to recombination emission) 
$j_D / j_{{\rm H}\alpha}\lesssim1$. Thus dust scattering will contribute about
the same amount to the peak halo surface brightness as recombination emission.

The ratio $j_D / j_{{\rm H}\alpha}$ has a radial dependence of 
$\propto (R / R_{\rm min})^{\eta - 2}$. For $\eta = 4$ this means that dust
scattering will dominate over recombination emission for increasing distances,
effectively lowering the steepness of the halo profile. This can be used to 
discriminate scattering from recombination emission. Such an attempt should be 
complemented by deep observations in [O III], as the Crab nebula is bright in
this line and dust scattering more efficient for shorter wavelengths.

\subsubsection{\label{collexcit}Collisional excitation}
One way to significantly enhance H$\alpha$ emission is collisional excitation 
of Ly$\beta$, followed by part of the radiative de-excitation going into 
H$\alpha$. This only works for temperatures well above $10^4$~K, agreeing with
our assumptions of $2\EE4$ K. A neutral fraction of hydrogen of 10\% is then
enough to increase H$\alpha$ by a factor of $\sim 10$ compared to pure 
recombination emission. If the shell gas was clumpy it is likely that such 
conditions prevail. However, our observations and those of \cite{fsh97} 
demand collisional excitation to be less dominant, as otherwise the halo would
have been detected unamiguously. It remains unclear though if this effect will
be constant with distance from the edge of the nebula, which could alter both
the halo's slope and amplitude. Detailed modelling is required to test this.

\section{\label{specresults}Prospects for a spectroscopic halo detection}
Direct imaging is not the only option for detecting a fast shell. \cite{fsh97}
took 9 ksec spectra with a 2.4m telescope of regions outside the nebula, but 
could not detect any emission lines. Such emission would only have small
Doppler shifts as the gas in this area moves mainly perpendicular to the line
of sight. Another way would be spectroscopy of objects near the explosion
centre, where the gas moves mainly along the line of sight and hence has a
very large Doppler shift. In this way \cite{sll00} detected blue-shifted C~IV
absorption which is currently the best observational evidence for a fast
shell.

Below we evaluate the chances of detecting the H$\alpha$ footprint of a shell 
towards the centre of the nebula. It would appear as emission on both sides of
the H$\alpha$ line, serving as an additional verification of an expanding
shell. Yet such an observation would be made against a complex and bright
H$\alpha$ background, and thus the location where the spectra are taken must
be picked carefully. A good choice will maximise the contrast as well as the
spectroscopic separation of the shell and the nebula emission.

\begin{figure}[t]
\center
\includegraphics[width=1.0\hsize]{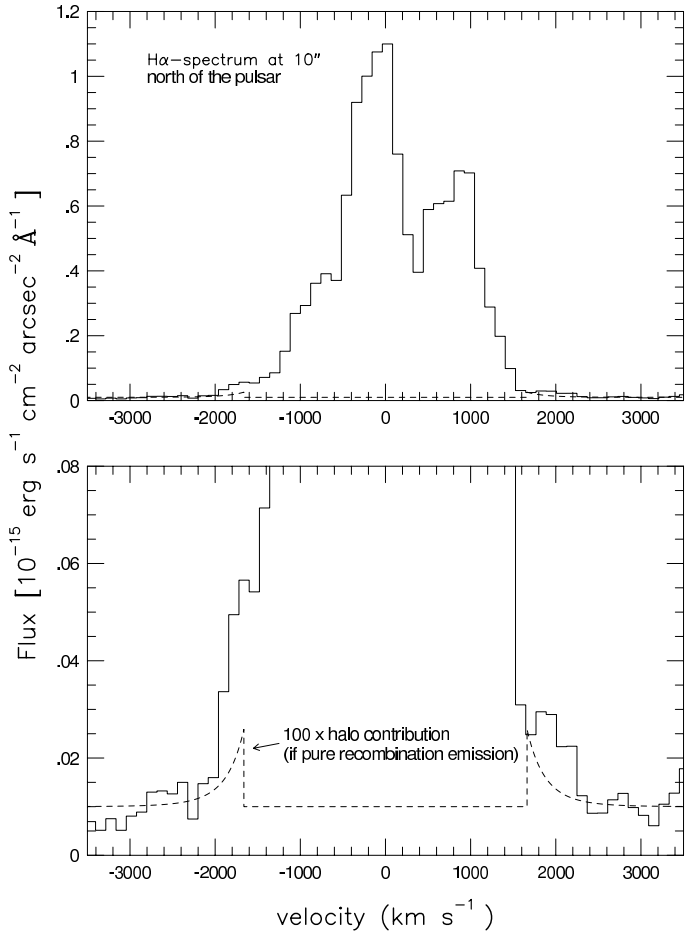}
\caption{Dereddened spectrum around H$\alpha$ 10\myarcsecnodot north of the
  pulsar. In the lower panel we have inflated the flux scale and added the
  $\eta = 4$ model spectrum from Fig. \ref{lineprofs_plot}. A constant of
  $10^{-17}\funit$ was included, simulating the continuum emission from the
  pulsar wind nebula. Note that the model spectrum has been multiplied by a
  factor of 100 to be distinguishable in this plot.}
\label{observedspectrum_plot}
\end{figure}

A set of FORS1 \citep{aff98} VLT spectra of the pulsar, taken for a different
programme, serves as our testbed. The total exposure time was 600 seconds
through a 1\myarcsec0 wide slit, the dispersion is 2.7 \angstromblank 
pixel$^{-1}$. We identified an area 10\myarcsecnodot north of the pulsar which
is largely free of bright filaments (see Fig.~\ref{m1neg}). The spectrum
around the H$\alpha$ line is shown in Fig. \ref{observedspectrum_plot}.

\begin{figure}[t]
\center
\includegraphics[width=1.0\hsize]{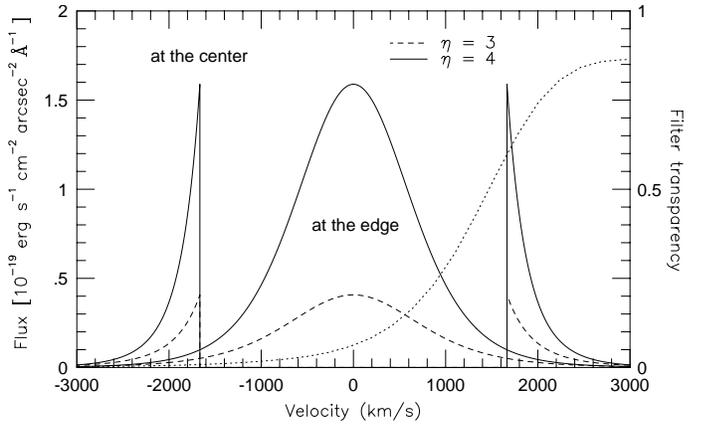}
\caption{Line profiles for the H$\alpha$ recombination emission of a fast
  shell, based on the models of Sect. \ref{shellmodels}. Note the gap in halo
  emission between $\pm 1680 \kms$ for lines of sight close to the explosion
  centre. The smaller the slope $\eta$ of the density profile, the less
  pronounced is the intensity fall off with increasing absolute velocity. The
  dotted line shows the filter curve of the WFI red continuum filter.}
\label{lineprofs_plot}
\end{figure}

We base our analysis on the shell models from Sect. \ref{shellmodels} and
their recombination emission. The H$\alpha$ line profile from such a shell 
depends strongly on the positioning of the spectroscopic slit, as gas with
different velocities is projected along different lines of sight. For example,
as the shell has an inner radius $R_{\rm min}$, no emission close to the
pulsar will be seen for velocities $|V|<1680\kms$ ($\pm37$~\AA) from the rest
wavelength of H$\alpha$. This gap decreases for lines of sight further away
from the centre. At the edge of the nebula the halo emission peaks at the rest
wavelength of H$\alpha$. The corresponding spectra are shown in
Fig. \ref{lineprofs_plot}.

In Fig. \ref{observedspectrum_plot} we overplotted the emission expected for
the $\eta=4$ model over the H$\alpha$ line observed. The model was increased
by a factor of 100 compared to the data in order to make it visible. Even if
collisional excitation was very important and boosted the halo's surface
brightness by a factor 100, the emission would still be undetectable in these
spectra. The main problem are the neighbouring [N~II] lines at 6548 \AA\ and
6583 \AA\, corresponding to $-675 \kms$ and $+944 \kms$, respectively. Their
rest wavelengths are close to the expected H$\alpha$ peak emissions of the
shell. The motions of the H$\alpha$ filaments with $\sim \pm1400 \kms$
contaminate the halo signal further. Increasing the integration times
significantly would not help to resolve this issue. We tested other areas
along the slit as well and they produce no better results than this one. We
therefore conclude that a search for H$\alpha$ shell emission inside the Crab
nebula is unfeasible.

\section{Discussion}
\subsection{\label{galhalos}PSF scattering and haloes around galaxies}
Our findings concerning the PSF scattering are applicable to other situations
where very faint haloes are searched for around bright extended targets. The
effect is easily overlooked, as the PSF is often thought of in scales of
arcseconds rather than of arcminutes. The PSF can be very efficient in
scattering light from far inside a bright target into its outskirts.

Observations of the extraplanar gas around edge-on galaxies are particularily
affected by this, as galaxies can have high central surface brightnesses and 
small minor axes. \cite{wbd02} recognised this effect in their work on NGC 
4565 and constructed large PSF models based on separate observations of
isolated bright stars. Their models extend to $28^{\prime}$ (corresponding to
1000 pixels), and they find that PSF scattering in their data is measurable
but negligible. \cite{wbd02} are cited by several authors working on similar
subjects, yet PSF scattering appears to be largely ignored.

\cite{zwb04} take this effect into account in their analysis of the stacked 
halo of 1047 edge-on galaxies taken from SDSS. They find a maximum
contamination of about 30\%. However, their PSF extends only to about
20 pixels ($8^{\prime\prime}$) and is then extrapolated by an exponential
function. \cite{dej08} investigated this in detail and reconstructed the SDSS
PSF out to 180 pixels. Beyond that, the PSF is extrapolated with a power law
index of $-2.6$ as obtained from the inner 40 to 180 pixels. Using a full 2D
convolution of the stacked galaxy image, \cite{dej08} shows that the PSF
contributes 50\% in $g$ and $r$, and up to 80\% in $i$-band to the signal
measured by \cite{zwb04}.

Even though this is an entirely different field of astrophysics, it suffers
from the same observational effect. In our case the situation is rather
extreme as the Crab nebula is of effectively constant surface brightness over 
its $5^{\prime}\times7^{\prime}$ area when smoothing out the filaments. Thus
PSF scattering is very effective as the entire object contributes. The
conclusions of \cite{dej08} are in excellent agreement with our own:
\begin{itemize}
\item{Extrapolating a PSF to large radii can be very misleading as a small
    change in the wing amplitude can have a large effect on the amount
    of scattered light.}
\item{The PSF wings should ideally be constructed from extremely bright 
    isolated stars and be joined with a core determined from bright but
    unsaturated stars.}
\item{The amount of scattering is very difficult to judge based just upon the
    PSF profile. Only a full 2D convolution will reveal its overall effect.}
\end{itemize}

\subsection{\label{perspectives}Prospects for future observations}
Our images are strongly affected by PSF contamination. This also applies 
for similar attempts at other telescopes, and can only be overcome by large
and accurate PSF models. These must be created by means of a very bright star
that by far dominates all other flux in the image. Ideally, the star
should be positioned on the same detector area as the target. It should be
bright enough so that the PSF can be modelled without using azimuthal
averaging, which destroys asymmetries in the PSF wings. Such a star should
ideally be observed close to the target and in the same night, bracketing the
target observations in time. The detector must be given sufficient time for
recovery after such observations so that charge persistencies can decay. Dark
conditions are required to keep the sky background as low as possible.

Observations through a red continuum filter would significantly minimise the 
contamination by scattered light. We investigated the efficiency of this
approach based on the WFI continuum filter. As can be seen from 
Figs. \ref{surface_halpha_plot} and \ref{lineprofs_plot}, the flux expected in
that filter would be 5 (8) times lower for the $\eta=3$ (4) models as compared
to H$\alpha$. This will be very difficult to record on 8m class telescopes. 
On the other hand, the combined emission from the processes discussed in Sect.
\ref{shellmodels}, or from its constituents alone, would become visible in
several hours H$\alpha$ exposure with e.g. FORS2 on the VLT, provided that
good PSF models are available. For an optimum comparison of such data with
model predictions one would subtract the 2D PSF-convolved image from the
original image, and then integrate over the halo in concentric rings around
the explosion centre.

On the spectroscopic side, the most promising approach is to search for 
Doppler shifted absorption features in a relatively uncontaminated spectral
region against the central part of the nebula. However, care has to be taken
in the interpretation of the result, as a chimney-like structure extending
towards us may contain gas at velocities similar to those in the shell. This
can be tested by searching for fast ($\sim2000\kms$) blueshifted [O~III] 
$\lambda 4959$ emission as [O~III] is bright in the Crab chimney. This would
lead to emission at around $4925-4935$ \AA\ and could be difficult to
disentangle from He I 4921. Adding information from other emission lines seen
in the chimney may be useful. 
%A possible way to improve on the results of \cite{sll00} would be to obtain 
%multi-object spectroscopy of point sources in the inner half of the Crab 
%nebula, looking for similar absorption features in possible background 
%sources.

Finally, deep 21 cm radio observations outside the nebula may be useful. The
main difficulty here is the uncertain degree of ionisation of hydrogen in the
shell. If collisional excitation of H-alpha is important, then a good fraction
of neutral hydrogen is implied which may show up in 21 cm line emission. Hence
observations of the H$\alpha$ surface brightness of the shell can help to
estimate the radio flux.

\section{Summary}
We have presented WFI and MOSCA images in H$\alpha$ to search for a high
velocity shell around the Crab nebula. Our detection limit in both data sets
is $5\EE{-8}\funit$, three times deeper than the previously deepest study by
\cite{fsh97}. However, our measurement of an H$\alpha$ halo is corrupted by
PSF contamination which can account for all of the signal. A real halo with
an observed (i.e. reddened) peak brightness of $2\EE{-7}\funit$ could
still be accomodated within the error bars. Thus we could not improve on the 
limits set by \cite{fsh97}. Deep imaging with 8m class telescopes, together 
with good PSF models, could yield progress in this field.

We discuss three different processes that could power H$\alpha$ emission from
a fast shell: recombination emission, dust scattering, and collisional
excitation. All but the latter yield flux levels about one order of magnitude
below the one observed. As the halo flux in two independent imaging data sets
is well described by PSF scattering, we can rule out collisional excitation to
play a dominant role in the shell gas.

We argue that a spectroscopic detection of H$\alpha$ emission towards the
centre of the nebula is unfeasible. The emission is very weak and buried in
the complex spectral environment of the H$\alpha$ and [N~II] lines. The
prospects of detecting blueshifted absorption features in the pulsar or
other background sources are significantly better and were already
demonstrated by \cite{sll00}. A deeper spectrum outside the Crab nebula could
still be very useful, as the recombination level is below the upper value from
~\cite{fsh97}. However, the amount of PSF scattering needs to be quantified
also for spectroscopy. This problem could be avoided by looking only at
emission further away than about $2200 \kms$ from H$\alpha$. However, the
flux expected for density profiles steeper than $\eta\sim2$ is so low that a
positive detection appears unfeasible.

\begin{acknowledgements}
We are grateful to Andrew Cardwell, Roger Chevalier and Rob Fesen for comments
on an early version of this manuscript. We thank G\"oran Olofsson for 
crucial discussions of scattered light, and the anonymous referee for very
useful suggestions. This work was supported by the Swedish Research
Council. At the start of this project PL was, and JS currently is, a Research
Fellow at the Royal Swedish Academy supported by a grant from the Wallenberg
Foundation.
\end{acknowledgements}

\bibliography{crab20090206}

\appendix
\section{MOSCA peculiarities}
We discovered that the gains of the individual chips in the MOSCA array depend
on the illumination level. The gain ratios are different by up to 20\% in flat
fields with approximately 30 kADU as compared to the observations with very
low background level ($\sim$ 100 ADU). The same behaviour was observed in data
taken through two other filters about two months later.

The THELI pipeline assumes that all chips of a mosaic camera are brought to
the same gain during flatfielding. This is not the case for MOSCA. We
circumvented the problem by switching off THELI's gain adjustment and 
processed and stacked the data from each CCD independently. The 4 stacked
images were then run again through the astrometric and photometric part of
THELI, being interpreted as 4 dithered exposures coming from a single-chip
camera. Given the overlap sources, relative photometric zeropoints could
be calculated and thus the gain differences were removed. Upon comparing the
fluxes of 74 common stars in the MOSCA and WFI data, we find a scattering of
0.08 mag without systematic trends as a function of position in the MOSCA
mosaic.

\end{document}